\documentclass[twocolumn]{autart}    

\usepackage{graphicx}          

\usepackage{amssymb}
\usepackage{amsmath}
\usepackage{xcolor}
\usepackage{soul}
\usepackage{booktabs}
\usepackage{multirow}
\usepackage{siunitx}
\usepackage{bm}
\usepackage[british,base]{babel}

\newcommand{\figref}[1]{Fig.~\ref{#1}}
\newcommand{\tabref}[1]{Table~\ref{#1}}
\newcommand{\inv}{^{-1}}
\newcommand{\tran}{^{\text{T}}}

\newtheorem{theorem}{Theorem}

\graphicspath{{images/pdf/}}

\begin{document}

\begin{frontmatter}

\title{VRFT with ARX controller model and \\ constrained total least squares\thanksref{footnoteinfo}} 

\thanks[footnoteinfo]{This study was financed in part by the Coordena{\c c}{\~ a}o de Aperfei{\c c}oamento de Pessoal de N{\'i}vel Superior - Brasil (CAPES) - Finance Code 001.
This work was financed in part by the Conselho Nacional de Desenvolvimento Científico e Tecnológico (CNPq).
The material in this paper will be partially presented at the 21st IFAC World Congress, July 12 - 17, 2020, Berlin, Germany.}

\author[Brazil]{Cristiane Silva Garcia}\ead{cristiane.garcia@ufrgs.br},    
\author[Brazil]{Alexandre Sanfelici Bazanella}\ead{bazanella@ufrgs.br}               

\address[Brazil]{Data-Driven Control Group, Department of Automation and Energy, Federal University of Rio Grande do Sul, Porto Alegre-RS, Brazil}  


\begin{abstract}                          
The virtual reference feedback tuning (VRFT) is a non-iterative data-driven (DD) method employed to tune a controller's parameters aiming to achieve a prescribed closed-loop performance.
In its most common formulation, the parameters of a linearly parametrized controller are estimated by solving a least squares (LS) problem, which in the presence of noise leads to a biased estimate of the controller's parameters.
To eliminate this bias, an instrumental variable (IV) variant of the method is usual, at the cost of increasing significantly the estimate's variance.
In the present work, we propose to apply the constrained total least squares (CTLS) solution to the VRFT problem. We formulate explicitly the VRFT solution with CTLS for controllers described by an autoregressive exogenous (ARX) model.
The effectiveness of the proposed solution is illustrated by two case studies in which it is compared to the usual VRFT solutions and to another, statistically efficient, design method.
\end{abstract}

\end{frontmatter}

\section{Introduction}
Virtual reference feedback tuning (VRFT) is a well-known and quite successful non-iterative data-driven (DD) method for control design.
It is based on the model reference paradigm, which defines a priori the desired closed-loop performance, that is, the reference model, and the controller's structure.
These informations are used, along with the data collected, to generate the (virtual) input-output signals employed in the identification of the best controller with the available controller structure \cite{campi2002virtual}.

Along the years, several works have extended or applied the VRFT method.
The case when the process presents non-minimum phase (NMP) zeros has been solved in \cite{campestrini2011virtual}.
The multivariable case is approached in \cite{nakamoto2004application} considering a diagonal parametrization of the reference model matrix, in \cite{campestrini2016unbiased} by dealing with fully parametrized reference models, and in
\cite{silva2017multivariable} multivariable NMP plants are considered.
Different applications of VRFT can be found in \cite{campi2003application,formentin2018robust,previdi2010virtual,rojas2012application,vanheusden2010identification}, for instance.

Although the VRFT method is not restricted to the linearly parametrized (LP) controllers case, this assumption is considered in its usual formulation and in most applications in the literature.
In fact, the controller chosen commonly has a fixed denominator and a parametrized numerator.
In the LP case, the solution is easily obtained by solving an ordinary least squares (OLS) problem.
Unfortunately, this solution produces a biased estimate in the presence of noise.
This bias is inherent to the problem's formulation, and it is expected even for a full-order controller, since it is due to the fact that the VRFT is not a standard identification problem.
The effect of the noise is usually counteracted by the application of an instrumental variable (IV), as suggested in \cite{campi2002virtual}.
However, the IV alternative has the drawback of increasing the variance of the estimate, which may lead to significant loss of performance and even to closed-loop instability \cite{soderstrom2006errors}.

With that in mind, in \cite{garcia2020constrained} we addressed the case of noisy data by applying a different solution to the VRFT problem: the constrained total least squares (CTLS). The CTLS solution removes the bias of the estimate and results in smaller variance, thus providing significant improvement in the closed-loop behaviour when compared to the IV solution.
In that work we considered the usual formulation of the VRFT, that is, the case of LP controllers with fixed poles.
In the present work we extend the analysis of \cite{garcia2020constrained} to the case of controllers with ARX structure, and provide further numerical evidence of the method's efficacy.
In addition of comparing the results obtained against the original solutions, we also compare against the results obtained with the optimal controller identification (OCI) method.
The OCI is a state-of-the-art DD method that is known to provide an efficient estimate of the controller's parameters.

The remaining of this paper is organized as follows.
Section~\ref{sec:preliminaries} shows some basic definitions.
The VRFT method with ARX controller structure is presented in section~\ref{sec:vrft}.
The CTLS, OLS and IV solutions are introduced in section~\ref{sec:solutions}.
The case studies are presented in Section~\ref{sec:simulations}.
Finally, section~\ref{sec:conclusions} shows the conclusions and future work.

\section{Preliminaries}
\label{sec:preliminaries}
Consider a linear discrete-time single-input single-output process
\begin{equation}
  y(t) = G(q)u(t) + H(q)v(t),
\label{eq:output_ol}
\end{equation}
where $G(q)$ is the process' transfer function, $H(q)$ is the noise model, $y(t)$ is the output signal, $u(t)$ is the input signal, $v(t)$ is zero-mean white noise with variance $\sigma^2$, and $q$ is the forward-shift operator.
$G(q)$ and $H(q)$ are assumed to be rational and causal transfer functions.

Considering data collected during an open-loop experiment, the input $u(t)$ is an exogenous signal, while the output signal $y(t)$ is affected by noise.
On the other hand, considering data collected during a closed-loop experiment, the input $u(t)$ is defined as
\begin{equation}
  u(t) = C_0(q) \left[r(t) - y(t) \right],
\label{eq:input_cl}
\end{equation}
where $C_0(q)$ is a stabilizing controller operating in the loop, and $r(t)$ is the reference signal.
Replacing \eqref{eq:input_cl} in \eqref{eq:output_ol} results in following model for the output signal:
\begin{equation}
  y(t) = T_0(q)r(t) + S_0(q)H(q)v(t),
\label{eq:output_cl}
\end{equation}
where $S_0(q) = (1+C_0(q)G(q))\inv$ is the original sensitivity transfer function and $T_0(q) = S_0(q)G(q)C_0(q)$ is the complimentary sensitivity transfer function.
Similarly, \eqref{eq:input_cl} is rewritten using \eqref{eq:output_ol} as
\begin{equation}
  u(t) = S_0(q) \left[C_0(q)r(t) - C_0(q)H(q)v(t) \right].
\label{eq:input_cl_final}
\end{equation}

\section{Virtual Reference Feedback Tuning with ARX controller}
\label{sec:vrft}
VRFT is a one shot DD method, which means that the controller's parameters can be tuned using a single batch of open- or closed-loop data.
Therefore, no special experiment is required, sufficing to have sufficiently rich data \cite{bazanella2011data}.
The VRFT method for ARX controllers is described in the following paragraphs.

First, let the controller $C(q, \bm\rho)$ be split into two components, as in \cite{campestrini2016data}:
\begin{equation}
  C(q,\bm\rho) = C_I(q,\bm\rho) C_F(q),
\label{eq:cont_split}
\end{equation}
where $C_F(q)$ is the fixed part, $C_I(q, \bm\rho)$ is the part to be identified, and $\bm\rho$ is the parameter vector.
This formulation is particularly convenient because it allows incorporating already-known information about the optimal controller into the controller's fixed part, such as a pole at $q=1$ to guarantee tracking constant references.

Now, define the reference model $M(q)$ representing the desired behaviour of the closed-loop $T(q,\bm\rho)$.
The method's objective is to find the parameter vector $\bm\rho$ that minimizes the reference tracking cost function
\begin{align}
  J_y(\bm\rho) &= \lim_{N\rightarrow\infty} \frac{1}{N} \sum\nolimits_{t=1}^N \left[ \left(T(q, \bm\rho) - M(q) \right)r(t) \right]^2
\label{eq:jy_rho} \\
  &= \lim_{N\rightarrow\infty} \frac{1}{N} \sum\nolimits_{t=1}^N \left[y(t, \bm\rho) - y_d(t) \right]^2,
\label{eq:jy}
\end{align}
where $y(t, \bm\rho)=T(q,\bm\rho)r(t)$ is the closed-loop output obtained with the controller $C(q, \bm\rho)$, and $y_d(t)=M(q)r(t)$ is the desired output.

The controller parametrization delimits a subspace within the controller space, known as the controller class:
\begin{equation*}
  \mathcal{C} = \left\lbrace C(q, \bm\rho) \,\ \vert \,\ \bm\rho \in \Omega \subseteq \mathbb{R}^{m} \right\rbrace,
\end{equation*}
where $\Omega$ is the subset of all implementable parameters and $m$ is the number of parameters.
Moreover, from \eqref{eq:jy_rho} and considering $T(q,\bm\rho) = (1+G(q)C(q,\bm\rho))\inv G(q)C(q,\bm\rho)$, the ideal controller could be calculated as
\begin{equation}
  C_d(q) = \left[G(q) - M(q)G(q) \right]\inv M(q),
\label{eq:ideal_controller_def}
\end{equation}
if $G(q)$ were available.
The ideal controller $C_d(q)$ is assumed to be in the controller's class $\mathcal{C}$, that is,
\[
  \exists \bm\rho_d \in \Omega \,\ \vert \,\ C(q, \bm\rho_d) = C_d(q),
\]
where $\bm\rho_d$ is the ideal parameter vector.

Pretending that the ideal controller $C_d(q)$ is already in the loop, one can generate the virtual reference as
\begin{equation}
  \bar{r}(t) = M\inv (t) y(t),
\label{eq:ref_virt}
\end{equation}
which is the signal that should be applied to the desired closed-loop to generate the output $y(t)$ collected.
After that, the virtual error $\bar{e}(t)$ is calculated as
\begin{align}
  \bar{e}(t) &= \bar{r}(t) - y(t) \nonumber \\
  &= \left(M\inv(q) - 1 \right)y(t) \label{eq:err_virt}.
\end{align}
where \eqref{eq:err_virt} was obtained using the definition of the virtual reference \eqref{eq:ref_virt}.

The virtual error is the input of the ideal controller, and the input of the part to be identified is given by \begin{align}
  \bar{e}_F(t) &= C_F(q)\bar{e}(t) \nonumber \\
  &= L_F(q)y(t), \label{eq:ef_def}
\end{align}
where \eqref{eq:ef_def} was obtained using \eqref{eq:err_virt}, and $L_F(q) = C_F(q) \left(M\inv(q) - 1 \right)$.
Possessing the input and output signals, $\bar{e}_F(t)$ and $u(t)$ respectively, finding the parameters of $C_I(q, \bm\rho)$ is a regular identification problem.
Therefore, the to-be-identified part is written as $C_I(q, \bm\rho) = B(q, \bm\rho)/A(q, \bm\rho)$, with
\begin{align}
  B(q,\bm\rho) &= b_1 + b_2q^{-1} + \dots + b_{n_b}q^{-n_b+1} \label{eq:b_def}\\
  A(q,\bm\rho) &= 1 + a_1q^{-1} + \dots + a_{n_a}q^{-n_a},\label{eq:a_def}
\end{align}
where the relative degree of $C_I(q, \bm\rho)$ is assumed zero, whereas $n_b$ and $n_a$ are the number of parameters of the numerator and denominator of $C_I(q,\bm\rho)$, respectively, i.e. $m=n_b+n_a$.

With the above definitions, we convert the VRFT problem into the problem of identifying the parameters of the following ARX model:
\begin{equation*}
  u(t) = B(q,\bm\rho)\bar{e}_F(t) + \left[1-A(q,\bm\rho) \right]u(t) +\varepsilon(t),
\end{equation*}
where $\varepsilon(t)$ is the model error.
The predictor $\hat{u}(t)$ is given by $\hat{u}(t) = B(q,\bm\rho)\bar{e}_F(t) + \left[1-A(q,\bm\rho) \right]u(t)$,
that can be rewritten using \eqref{eq:b_def} and \eqref{eq:a_def} as
\begin{align}
  \hat{u}(t) &= \bm\rho\tran \bm\varphi(t), \label{eq:u_model}
\end{align}
where the parameter vector $\bm\rho$ and the regressor vector $\bm\varphi(t)$ are given by
\begin{align}
  \bm\rho\tran &= [b_1 \, b_2 \, \dots \, b_{n_b} \, a_1 \, a_2 \, \dots \, a_{n_a}] \nonumber \\
  \bm\varphi(t) &= \left[\bar{e}_F(t) \, \bar{e}_F(t-1) \, \dots \, \bar{e}_F(t-n_b+1) \right. \nonumber \\
  & \quad \left. -u(t-1) \, -u(t-2) \, \dots  \, -u(t-n_a) \right]\tran.
\label{eq:phi_arx}
\end{align}
Now, using the above information, the VRFT transforms the problem of minimizing the cost function $J_y(\bm\rho)$ presented in \eqref{eq:jy_rho} into a least squares identification of the controller $C_I(q,\bm\rho)$, which consists in minimizing the following cost function
\begin{equation*}
  J^{\text{VR}}(\bm\rho) = \lim_{N\rightarrow\infty} \frac{1}{N} \sum\nolimits_{t=1}^N \left[u(t)-\bm\rho\tran\bm\varphi(t) \right]^2.
\end{equation*}
Assuming that the system is not affected by noise and that there is an ideal controller $C(q,\bm\rho_d) = C_I(q,\bm\rho_d)C_F(q)$ such that $J_y(\bm\rho_d) = 0$, that is $T (q,\bm\rho_d) = M(q)$, the minimum of $J_y(\bm\rho)$ is proven to coincide with the minimum of $J^{\text{VR}}(\bm\rho)$.
Note that it is the same cost function of the original formulation of the VRFT method.
The difference between the usual and the ARX formulations reside in how the regressor vector $\bm\varphi(t)$ is constructed.
Considering the ARX structure, the regressor vector in \eqref{eq:phi_arx} is formed using $\bar{e}_F(t)$ and $u(t)$.
However, in the usual formulation $u(t)$ does not appear in the regressor vector, because the controller structure is assumed to have a fixed denominator.

\section{Solutions for the VRFT problem}
\label{sec:solutions}
First, consider the noise-free case, and let the subscript $_0$ denote that a quantity has been generated in this situation; then we can write
\begin{equation}
  \bm u_0 = \bm\Phi_0 \bm\rho_d,
\label{eq:noise_free_problem}
\end{equation}
where $\bm\rho_d$ is the actual (desired) parameter vector, $\bm u_0 \in \mathbb{R}^N$ is the controller's output vector given by $\bm u_0 = [u_0(1)\,u_0(2)\,\dots \,u_0(N)]\tran$, $\bm\Phi_0 \in \mathbb{R}^{N\times m}$ is the noiseless regressor matrix defined as $\bm\Phi_0 = [\bm\varphi_0(1)\, \bm\varphi_0(2)\, \dots \, \bm\varphi_0(N)]\tran$ and $N$ is the number of samples.

In the more realistic case of data affected by noise, the calculated regressor matrix $\bm\Phi$ and the measured output vector $\bm u$ may be written as
\begin{align}
  \bm\Phi &= \bm\Phi_0 + \bm\Delta_\Phi \label{eq:phi}\\
  \bm u &= \bm u_0 + \bm\delta_u \label{eq:u}
\end{align}
where $\bm u = \left[u(1) \,\ u(2) \,\ \dots \,\ u(N) \right]\tran$ and $\bm\Phi = \left[ \bm\Phi_e \,\ \bm\Phi_u \right]$.
Here, $\bm\Phi_e \in \mathbb{R}^{N\times n_b}$ and $\bm\Phi_u \in \mathbb{R}^{N\times n_a}$ are defined as
\begin{align}
  \bm\Phi_e &=
  \begin{bmatrix}
    \bar{e}_F(1)  & 0               & \dots   & 0                   \\
    \bar{e}_F(2)  & \bar{e}_F(1)    & \dots   & 0                   \\
    \vdots        & \vdots          & \ddots  & \vdots              \\
    \bar{e}_F(N)  & \bar{e}_F(N-1)  & \dots   & \bar{e}_F(N-n_b+1)
  \end{bmatrix} \label{eq:phi_e}\\
    \bm\Phi_u &=
    \begin{bmatrix}
      0       & 0       & \dots   & 0 \\
      -u(1) & 0       & \dots   & 0 \\
      \vdots  & \vdots  & \ddots  & \vdots \\
      -u(N-1) & -u(N-2) & \dots   & -u(N-n_a)
    \end{bmatrix}. \label{eq:phi_u}
\end{align}
Moreover, $\bm\Delta_\Phi \in \mathbb{R}^{N \times m}$ and $\bm\delta_u \in \mathbb{R}^N$ represent the noise contributions in $\bm\Phi$ and $\bm u$, respectively.
Isolating $\bm\Phi_0$ and $\bm u_0$ in \eqref{eq:phi} and \eqref{eq:u}, respectively, and replacing those in \eqref{eq:noise_free_problem} gives a general formulation for the problem of
estimating the parameter vector $\bm\rho$:
\begin{equation}
  \left(\bm\Phi - \bm\Delta_\Phi \right)\bm\rho = \bm u - \bm\delta_u.
\label{eq:gen_problem}
\end{equation}
In the next subsections three different solutions for this problem are presented: the CTLS, the OLS, and the IV solutions.

\subsection{The Constrained Total Least Squares solution}
\label{subsec:ctls}
Originated from the total least squares problem, the CTLS has as \emph{constraint} that the calculated regressor matrix $\bm\Phi$ and the collected output vector $u$ are affected by the same noise source \cite{abatzoglou1987constrained,abatzoglou1991constrained}.
This problem is equivalent to the structured total least squares as demonstrated in \cite{lemmerling1996equivalence}.

The CTLS problem is formulated from \eqref{eq:gen_problem} as
\begin{align*}
  \min_{\bm\rho, \bm v} \left\lVert \left[ \bm\Delta_\Phi \,\ \bm\delta_u \right] \right\rVert_\text{F}^2  \nonumber \,
  & \quad \text{{s. t.}} \,\ \left(\bm\Phi - \bm\Delta_\Phi \right)\bm\rho = \bm u-\bm\delta_u, \, \text{and}  \nonumber\\
  & {} \quad \left[ \bm\Delta_\Phi \,\ \bm\delta_u \right] = \left[\bm P_1 \bm v \,\ \bm P_2 \bm v \,\ \dots \,\ \bm P_{m+1} \bm v \right] \nonumber
\end{align*}
where $\lVert \cdot \rVert_\text{F}$ is the Frobenius norm, that is, $\lVert \bm X \rVert_\text{F}^2 = \sum_{i,j}\vert x_{i,j} \vert^2$, whereas $\bm v \in \mathbb{R}^N$ is a vector with the noise samples, that is, $\bm v = \left[v(1) \,\ v(2) \,\ \dots \,\ v(N) \right]\tran$,
and $\bm P_i \in \mathbb{R}^{N\times N}$ with $i=1,\dots,m+1$ are Toeplitz matrices representing filters.
These filters describe how the single noise source $\bm v$ affects every column of the matrix $\left[ \bm\Delta_\Phi \,\ \bm\delta_u \right]$.
Therefore, the CTLS problem minimizes the Frobebius norm of the noise contributions.
This problem was simplified in \cite{abatzoglou1987constrained} to remove the decision variable $\bm v$, and the solution is reproduced below.
\begin{theorem}
  Let $[\bm\Delta_\Phi \,\ \bm\delta_u] = [\bm P_1 \bm v \,\ \bm P_2 \bm v \,\ \dots \,\ \bm P_{m+1} \bm v]$.
  Then the estimated parameter vector $\hat{\bm\rho}$ is the solution of the following minimization problem
  \begin{align*}
    \min_{\bm\rho}
    \begin{bmatrix}
      \bm\rho \\ -1
    \end{bmatrix}\tran
    \begin{bmatrix}
      \bm\Phi & \bm u
    \end{bmatrix} \tran
    \left(\bm\Gamma_\rho \bm K\inv \bm\Gamma_\rho\tran \right)\inv
    \begin{bmatrix}
      \bm\Phi & \bm u
    \end{bmatrix}
    \begin{bmatrix}
      \bm\rho \\ -1
    \end{bmatrix},
  \end{align*}
where $\bm\Gamma_\rho = \sum\nolimits_{i=1}^m \left(\bm P_i \bm\rho_i\right) - \bm P_{m+1}$, and $\bm K = \sum\nolimits_{i=1}^{m+1} \bm P_i\tran \bm P_i$.
\label{theo:minimization_problem}
\end{theorem}
The proof is also given in \cite{abatzoglou1987constrained}.
To apply the CTLS solution to the VRFT problem, it is necessary to define the filters' impulse responses and to generate the matrices $\bm P_i$ with $i=1,2,\dots,m+1$ that will be used in the optimization.
These filters are different depending on the type of the experiment, and are described below for each case.

\textbf{Filters for the closed-loop case:}
In this case, both signals $y(t)$ and $u(t)$ are affected by noise, as presented in \eqref{eq:output_cl} and \eqref{eq:input_cl_final}, respectively.
The filters for the matrix $\bm\Delta_\Phi$ are obtained using \eqref{eq:phi_e} and \eqref{eq:phi_u}.
Observe that the columns of the matrix $\bm\Phi_e$, in \eqref{eq:phi_e}, are composed by the vector
\[
  \bar{\bm e}_F = [\bar{e}_F(1) \, \bar{e}_F(2) \, \dots \, \bar{e}_F(N)]\tran
\]
and its delayed versions.
In the same way, the columns of the matrix $\bm\Phi_u$, in \eqref{eq:phi_u}, are formed by the vector containing the input samples
\[
  \bm u = [-u(1)\, -u(2)\, \dots \, -u(N)]\tran
\]
and its delayed versions.
The to-be-identified controller's input, $\bar{e}_F(t)$, is rewritten replacing \eqref{eq:output_cl} in \eqref{eq:ef_def} as
\begin{equation}
  \bar{e}_F(t) = \underbrace{L_F(q)T_0(q)r(t)}_{\bar{e}_{F0}(t)} + \underbrace{L_F(q)S_0(q)H(q)v(t)}_{\bm\delta_{\bar{e}_{F}(t)}},
\label{eq:ef_split}
\end{equation}
where the signal is split into two terms: one purely from the reference signal, $\bar{e}_{F0}(t)$; and another with the noise contribution alone, $\bm\delta_{\bar{e}_{F}(t)}$.

In the same way, let the input signal \eqref{eq:input_cl_final} be rewritten as
\begin{equation}
  u(t) = \underbrace{S_0(q)C_0(q)r(t)}_{u_0(t)} - \underbrace{S_0(q)C_0(q)H(q)v(t)}_{\bm\delta_u(t)},
\label{eq:u_split}
\end{equation}
where the signal is split into two terms: one from the reference signal, $u_0(t)$, and another with the noise contribution, $\bm\delta_u(t)$.
Notice from \eqref{eq:ef_split} and \eqref{eq:u_split} that there is a common term $S_0(q)H(q)$ in the filters that generate $\bm\delta_{\bar{e}_{F}(t)}$ and $\bm\delta_u(t)$.
Because the common (coloured) noise source $S_0(q)H(q)v(t)$ affects every column of the matrix $[\bm\Delta_\Phi \,\ \bm\delta_u]$, the filters for the closed-loop case may be simplified to:
\begin{align}
  F_i(q) = \left\lbrace
  \begin{aligned}
    -L_F(q)q^{-(i-1)},   & \,\ i=1,\dots,n_b \\
    -C_0(q)q^{-(i-n_b)}, & \,\ i=n_b+1,\dots,m \\
    C_0(q),              & \,\ i=m+1
  \end{aligned}
  \right.
\label{eq:filters_cl}
\end{align}
Notice that those filters now depend only on information already known: the filter $L_F(q)$ and the controller operating in closed-loop $C_0(q)$.

\textbf{Filters for the open-loop case:}
In this scenario, only the output signal $y(t)$ is affected by noise, as described in \eqref{eq:output_ol}.
Because of that, only $\bar{e}_F(q)$ has a noise contribution, and consequently, the filters are obtained looking at the matrix $\bm\Phi_e$.
Replacing \eqref{eq:output_ol} in \eqref{eq:ef_def} gives
\begin{equation*}
  \bar{e}_F(t) = \underbrace{L_F(q)G(q)u(t)}_{\bar{e}_{F0}(t)} + \underbrace{L_F(q)H(q)v(t)}_{\bm\delta_{\bar{e}_F(t)}},
\end{equation*}
where, as before, this signal is split into two terms: one purely from the input signal, $\bar{e}_{F0}(t)$; and another with the noise contribution, $\bm\delta_{\bar{e}_F(t)}$.

There are $n_b$ filters and the remaining filters are zero, since the input signal $u(t)$ is not affected by noise.
Again, in order to simplify the filters, notice that $H(q)v(t)$ is the common (coloured) noise source affecting the columns of the matrix $[\bm\Delta_\Phi \,\ \bm\delta_u]$ and let the filters for the open-loop case be:
\begin{align}
  F_i(q) = \left\lbrace
  \begin{aligned}
    -L_F(q)q^{-(i-1)},   & \,\ i=1,\dots,n_b \\
    0, & \,\ i=n_b+1,\dots,m+1
  \end{aligned}
  \right.
\label{eq:filters_ol}
\end{align}
Observe that, also in this case, the filters depend only on the known filter $L_F(q)$.

\subsection{The original solutions}
\label{subsec:original}
VRFT's original solution, OLS, considers that only the output vector $u$ is corrupted by noise, this way, the problem is formulated considering $\bm\Delta_\Phi$ equals zero in \eqref{eq:gen_problem}.
The optimization problem is formulated as
\begin{equation*}
  \min_{\bm\rho,\bm\delta_u} \,\ \lVert\bm\delta_u \rVert^2 \quad \text{s. t.} \,\ \bm\Phi \bm\rho = \bm u-\bm\delta_u.
\end{equation*}
Observe that the minimization is w.r.t. $\bm\delta_u$ instead of $\bm v$, that is because the OLS disregards  the filter information presented before.
Nonetheless, the algebraic solution is given by
\begin{equation*}
  \hat{\bm\rho} = \left(\bm\Phi\tran \bm\Phi \right)\inv \bm\Phi\tran \bm u.
\end{equation*}
Following standard identification theory \cite{ljung1999system}, this solution leads to a biased estimate, because the regressor matrix $\bm\Phi$ is affected by noise.

As suggested in \cite{campi2002virtual}, an IV can be used aiming to eliminate the bias.
This paper considers the IV that requires performing a second experiment in the process with the same input signal to collect more data.
From those data, the pairs of signals $u_1(t)$, $y_1(t)$, $u_2(t)$ and $y_2(t)$ are generated and the parameters are estimated as
\begin{equation*}
  \hat{\bm\rho} = \left(\bm\Phi_1\tran \bm\Phi_2 \right)\inv \bm\Phi_1\tran \bm u_2,
\end{equation*}
where $\bm\Phi_1$ is generated using $y_1(t)$ and $u_1(t)$ in \eqref{eq:ef_def}, \eqref{eq:phi_e}, and \eqref{eq:phi_u}.
The same is applied to generate $\bm\Phi_2$, using $y_2(t)$ and $u_2(t)$.
Also, keeping the notation, $\bm u_2$ is a vector containing the samples of $u_2(t)$.
Although the IV approach leads to unbiased estimate it has the inconvenient of increasing its variance.
This increase in the variance may lead to a poor closed-loop performance and even closed-loop instability \cite{soderstrom2006errors}.
In fact, this effect is seen in the simulation results presented in the next section.

\section{Simulation examples}
\label{sec:simulations}
In order to evaluate the proposed solution, two simulation case studies are carried out along the lines of the examples in \cite{campestrini2016data}.
In all cases, 100 Monte Carlo simulations are performed varying the noise realization.
The process transfer function $G(q)$ is
\begin{equation}
  G(q) = \frac{0.5 (q-0.8)}{(q-0.7)(q-0.9)}.
\label{eq:process_exp}
\end{equation}
The reference model is given by
\begin{equation}
  M(q) = \frac{0.16q}{(q-0.6)^2}.
\label{eq:ref_model_exp}
\end{equation}
Replacing \eqref{eq:process_exp} and \eqref{eq:ref_model_exp} in \eqref{eq:ideal_controller_def} gives the following ideal controller:
\begin{equation*}
  C_d(q) = \underbrace{\frac{0.32(q-0.7)(q-0.9)}{(q-0.36)(q-0.8)}}_{C_I(q,\bm\rho_d)}\underbrace{\frac{q}{(q-1)}}_{C_F(q)},
\end{equation*}
where the fixed part contains an integrator and a zero at the origin, leaving five parameters to be estimated.
Regarding the collected output signal, a coloured noise corrupts $y(t)$.
This noise is generated by filtering a white noise sequence, with variance $\sigma^2$, through the following noise model
\begin{equation}
  H(q) = \frac{q}{q-0.3},
\label{eq:noise_tf}
\end{equation}
where the $\sigma^2$ varies depending on the experiment.

Two metrics are used to assess the design results: an estimate of the performance criterion cost function $\hat{J}_y(\hat{\bm\rho})$, and the parameters' Mean Squared Error (MSE).
An estimation of the cost function $\hat{J}_y(\hat{\bm\rho})$ is calculated for each controller, estimated in each Monte Carlo simulation, as
\[
  \hat{J}_y(\hat{\bm\rho}) = \frac{1}{N} \sum\nolimits_{t=1}^N \left[y(t, \hat{\bm\rho}) - y_d(t) \right]^2,
\]
where $N=100$.
The $\text{MSE}(\hat{\bm\rho})$ is given by
\begin{align}
  \text{MSE}(\hat{\bm\rho}) &= \operatorname{E} \lVert \hat{\bm\rho} - \bm\rho_d \rVert^2  \nonumber \\
  &\approx \frac{1}{N_{\text{sim}}} \sum\nolimits_{i=1}^{N_{\text{sim}}} \lVert \hat{\bm\rho}_i - \bm\rho_d \rVert^2, \label{eq:mse}
\end{align}
where $N_{\text{sim}} = 100$.

In each Monte Carlo simulation, the parameters are estimated using: the proposed solution as presented in subsection~\ref{subsec:ctls}; the OLS and IV solutions as described in subsection~\ref{subsec:original}; and the OCI method \cite{campestrini2016data}.
The CTLS optimization problem is non-convex, requiring an iterative algorithm and an initial guess.
In these case studies the Matlab's \verb|fminsearch| function was employed starting at $\bm\rho_0=0.8\bm\rho_d$.
Observe that this is just an illustrative example and other choices of algorithms and initializations are possible.
In practical situations, the estimates obtained from the IV or the OLS methods may be considered as initial guesses for the optimization.
Keep in mind that such choices should be considered carefully, because they may cause the optimization to converge to other local minima.
Regarding the OCI method, the Matlab's \verb|ident| toolbox is used to estimate the parameters, as suggested in \cite{campestrini2016data}.
Since the OCI is also non-convex, the same initial guess $\bm\rho_0=0.8\bm\rho_d$ is used to initialise the method.
The next subsections present the results obtained with data collected from open- and closed-loop simulated experiments.

\subsection{Open-loop data}
In the first case study, the parameters are estimated using data collected from an open-loop experiment, where a pseudorandom binary sequence (PRBS) with 1000 samples is applied as input signal $u(t)$.
The output signal $y(t)$ is affected by a white noise sequence with variance $\sigma^2 = 0.01$ filtered by \eqref{eq:noise_tf}.

\begin{figure}[b]
  \centering%
  \includegraphics{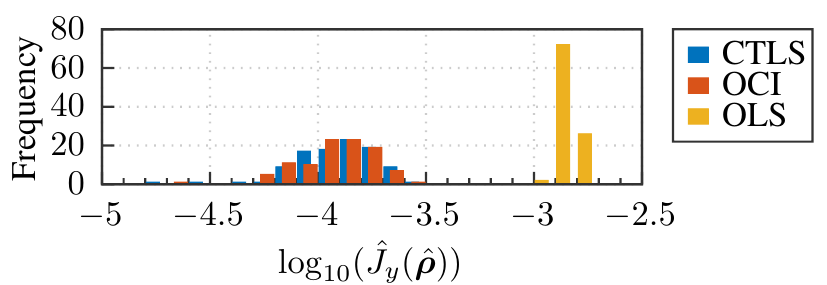}%
  \caption{\label{fig:histogram_ma_cf2} Histograms of $\log_{10}(\hat{J}_y(\hat{\bm\rho}))$ for the open-loop case.}
\end{figure}

\begin{figure}[b]
  \centering%
  \includegraphics{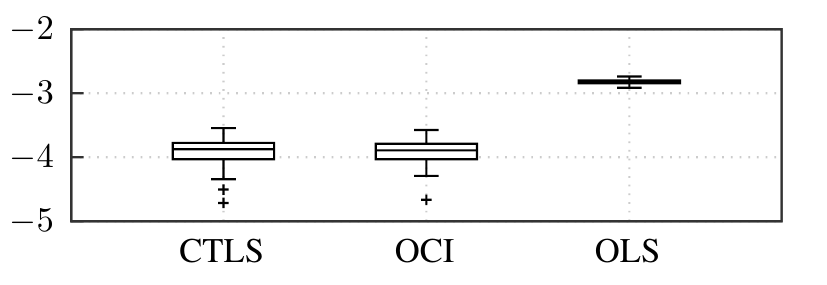}%
  \caption{\label{fig:boxplot_ma} Box plot of $\log_{10}(\hat{J}_y(\hat{\bm\rho}))$ for the open-loop case.}
\end{figure}

\figref{fig:histogram_ma_cf2} presents the histograms of the estimated cost functions obtained using each method.
Observe that the values obtained with the original OLS approach are much larger than the ones obtained with the proposed solution and with the OCI method.
That is expected, since the OLS estimate is known to be biased in the presence of noise.
In fact, the CTLS and OCI approaches give results approximately ten times better than the ones obtained with the OLS solution.
The results obtained with the IV approach are not presented here because only 52\% of the controllers resulted in stable closed-loop.
\figref{fig:boxplot_ma} presents the box plot of the $\log_{10}$ of the estimated cost function $\hat{J}_y(\hat{\bm\rho})$.
Note that the CTLS and OCI approaches give similar results, while the OLS solution gives the largest result.

The parameters' MSE, calculated as in \eqref{eq:mse}, are presented in the respective rows in \tabref{tab:mse}, along with the bias and variance of the parameters.
As expected, the OLS approach gives an estimation with large bias and small variance.
Also, both CTLS and OCI estimates present a great reduction in the bias, and even though the variance of the estimate is more pronounced, when compared with the OLS, the resulting MSE is significantly reduced.
Once again, because of the amount of controllers leading to unstable closed-loop responses, the results obtained with the IV are omitted.

\begin{table}[t]
  \caption{Values of the parameters' MSE\label{tab:mse}}
  \centering
  \begin{tabular}{llllllll}
  \toprule
   & Method & bias & variance & $\text{MSE}(\hat{\bm\rho})$ \\
  \midrule
  \multirow{3}{*}{Open-loop}
  & OLS   & 2.0995 & 0.0007 & 2.1002 \\
  & CTLS  & 0.0005 & 0.0076 & 0.0081 \\
  & OCI   & 0.0004 & 0.0059 & 0.0063 \\
  \midrule
  \multirow{3}{*}{Closed-loop}
  & OLS   & 2.0781 & 0.0007 & 2.0787 \\
  & CTLS  & 0.0003 & 0.0075 & 0.0077 \\
  & OCI   & 0.0003 & 0.0058 & 0.0060 \\
  \bottomrule
  \end{tabular}
\end{table}

\subsection{Closed-loop data}

In order to collect closed-loop data, other 100 experiments are also performed, where, instead of exciting $u(t)$ directly, the reference input $r(t)$ is excited by a PRBS with 1000 samples.
During the experiment a coloured noise is injected into the output.
This signal is generated by a white noise sequence with variance $\sigma^2 = 9 \times 10^{-4}$ filtered through \eqref{eq:noise_tf}.
The controller operating initially in closed-loop is given by
\[
  C_0(q) = \frac{0.3(q-0.7)(q-0.9)}{(q-0.8)(q-1)}.
\]

\begin{figure}[b]
  \centering%
  \includegraphics{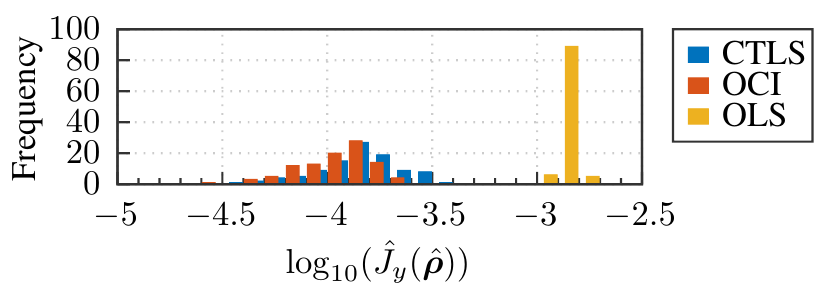}%
  \caption{\label{fig:histogram_mf_cf2} Histograms of $\log_{10}(\hat{J}_y(\hat{\bm\rho}))$ for the closed-loop case.}
\end{figure}

\figref{fig:histogram_mf_cf2} presents the histograms of the estimated cost functions calculated for each controller and each method.
Once again the values values obtained with the CTLS and OCI approaches are approximately ten times smaller than the ones obtained with the OLS.

\begin{figure}[t]
  \centering%
  \includegraphics{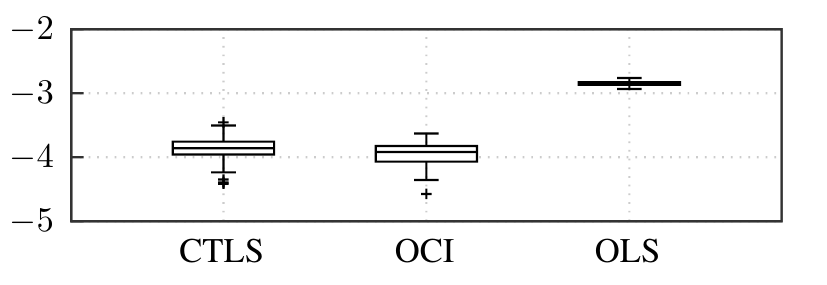}%
  \caption{\label{fig:boxplot_mf} Box plot of $\log_{10}(\hat{J}_y(\hat{\bm\rho}))$ for the closed-loop case.}
\end{figure}

\figref{fig:boxplot_mf} presents the box plot of the estimated cost function $\hat{J}_y(\hat{\bm\rho})$.
Observe that both CTLS and OCI approaches presented similar results.
On the other hand, the largest value is found with the OLS solution.

\tabref{tab:mse} presents the parameters' MSE calculated for each method in its respective rows.
In the same table, the values obtained for the bias and variance of the parameters are presented.
Observe that the OLS solution presents the largest bias and MSE values.
The CTLS and OCI present the smallest bias, but a variance more pronounced than the OLS solution.
As before, the results obtained with the IV approach are not presented, because 19\% of the controllers resulted in unstable closed-loop behaviour.

\section{Conclusions and future work}
\label{sec:conclusions}

In this work, the VRFT method is formulated considering a controller represented with an ARX structure.
This choice gives more flexibility than the choices commonly found in the literature.
Besides, we proposed a formulation that allows fixing part of the controller, if known a priori, reducing the number of parameters that need to be estimated.

In order to deal with output measurement noise, we proposed to apply the CTLS solution to estimate the controller's parameters, instead of the original (OLS and IV) solutions.
The feasibility of the proposed solution is presented in two simulation case studies, where the results are compared with the original solutions, and with the OCI method, which is unbiased and statistically efficient.
A comparison between the results shows that the CTLS gives better results compared with the original solutions, and similar results when compared with the OCI method.
Although the OCI method gives similar results to the CTLS, the latter has the advantage of not needing to identify the noise model, whose structure is usually unknown.
Besides, the number of parameters to be identified is smaller.

As future work, we intent to deal with the mismatched case, in which the ideal controller is not in the controllers' class.
These usually happens when the controller is underparametrized.
Also, we intend to extend the proposed solution to multivariable processes and non-minimum-phase processes.

\bibliographystyle{plain}        





\end{document}